\documentclass[showpacs]{revtex4}
\usepackage{epsfig}
\usepackage{amssymb}
\usepackage{amsmath}
\usepackage{graphicx}
\newcommand{\be}{\begin{equation}}
\newcommand{\ee}{\end{equation}}
\newcommand{\ben}{\begin{eqnarray}}
\newcommand{\een}{\end{eqnarray}}
\newcommand{\lb}{\label}
\begin{document}

\title{\bf Exotic Behavior of Heavy-Flavored Meson Matter}
\author{L. M. Abreu$^{1}${\footnote{email: luciano.abreu@ufba.br}},
E. S. Nery$^{1}${\footnote{email: elenilsonnery@hotmail.com}}    
and R. Rodrigues da Silva$^{2}${\footnote{email: romulo@df.ufcg.edu.br}}
}
\affiliation{$^{1}$Instituto de F{\'i}sica, Universidade Federal da
Bahia, 40210-340, Salvador-BA, Brazil}
\affiliation{$^{2}$
Unidade Acad\^emica de F{\'i}sica, Universidade Federal de Campina Grande, Campina Grande-PB, Brazil.}

\begin{abstract}

In this work, we study the thermodynamic behavior of heavy-flavored meson matter in the framework of $(\sigma,\omega)$-meson-exchange model
in relativistic mean field theory.
We find a decreasing of the effective masses of $D$ 
and $B$ mesons as the temperature increases. 
By using the effective mass and maximum value of dissociation temperatures available from lattice QCD, the masses of the bound states $D \bar{D}$ and $B \bar{B}$ are estimated in 2 MeV for both molecules. For the $B$-meson matter, the pressure presents an exotic behavior, being negative for temperatures above 6.6 times the deconfinement transition temperature $T_c$. In addition, the ratio of pressure to energy density is similar 
to the value predicted for systems that behave as dark energy matter.
 
\end{abstract}
\maketitle

\section{Introduction}

In 2003 the Belle Collaboration \cite{Choi:2003ue} reported the first experimental evidence for 
non-$q\bar{q}$ mesons. Thenceforth several new states (called $X$, $Y$ and $Z$) have been discovered in $B$-meson decays \cite{Choi:2003ue,Abe:2004zs,Aaij:2011sn,Choi:2007wga,Aaltonen:2009tz},
with masses between 3.9 GeV and 4.7 GeV for charmonium sector and $Z_b(10610)$ and $Z_b(10650)$
for bottomonium sector. 
These states, however, do not fit as excited states of charmonium 
described in the framework of constituent quark model \cite{Godfrey:1985xj}. 
Furthermore, they have an exotic decay modes,  in which charmed mesons are suppressed. 

Thus, a large amount of effort has been directed to understand the structure of these exotic mesons. In particular, one new meson that has strong evidence of its existence is the $X(3872)$, 
which has recently been measured in LHCb \cite{Aaij:2011sn}. The quantum numbers
that have been suggested by PDG \cite{Beringer:1900zz} are isospin $I=0$ and $J^{PC}=1^{++}$ or $J=2^{-+}$. From a theoretical point of view, several proposals have been attempted in order to provide some insight in this issue, as models based on glueball \cite{Seth:2004zb,Morningstar:1999rf}, 
mixed state $(c \bar{c}-D\bar{D}^{*})$  \cite{Barnes:2003vb}, tetraquark state \cite{Maiani:2004vq} and molecular state $(D^{0}\bar{D}^{*0} +  c.c.)$ \cite{Tornqvist:1993ng}. 

Notice that the notion of hadronic molecular states is not new, being proposed several decades ago \cite{Voloshin:1976ap,Novikov:1977dq}. Nevertheless, it has been largely employed to get some understanding about the properties of exotic states, mainly when the observed meson masses are rather close to a respective two-meson threshold. In the example discussed above, the mass of $X(3872)$ is very close to $D\bar{D}^{*}$-state threshold. The same idea is used to other exotic states, with possible molecules like $D^{*}\bar{D}^{*}$, $B\bar{B}^{*}$, $B^{*}\bar{B}^{*}$ and so on.

Pursuing in the molecule approach, another interesting possible combination of heavy-flavor mesons is a bound state between $D$ and $\bar{D}$, called $X(3700)$.
This state has spin-0 and isospin-0 and was predicted by Gamermann et al. \cite{Gamermann:2006nm}, 
but still here there is no experimental evidence for this state.
However, BES Collaboration \cite{Ablikim:2006zq, Ma:2008nj} 
observed in the process $e^{+}e^{-}\rightarrow D\bar{D}$ 
two states: the ``expected'' $\psi(3770)$ and 
a probable new state in the energy region 
between 3.700 and 3.872 GeV. In addition the results 
from branching fraction for $\psi(3770) \rightarrow D\bar{D}$
cannot be explained if $\psi(3770)$ is a pure $c\bar{c}$ state,
but if it contains four-quark admixture \cite{Voloshin:2005sd,Rong:2010ai}.

Many works have been devoted to study of bound state of $D$ and $\bar{D}$ that support the existence of $X(3700)$ \cite{Gamermann:2006nm,Ke:2012gm,Liu:2010xh,Li:2012mq,Zhang:2009vs}. 
In these analyses different frameworks were used, as the couple channels \cite{Gamermann:2006nm}, 
Bethe Salpeter equation \cite{Ke:2012gm}, 
meson exchange model \cite{Liu:2010xh}, chiral quark model \cite{Li:2012mq}, 
QCD sum rules \cite{Zhang:2009vs} and mean field approach \cite{Freire:2010zd}. However, it is worth remarking that there is in literature divergent understanding of $X(3700)$ as a molecular state \cite{Yang:2011rp}.
Also, it deserves mention the meson exchange approach discussed in 
Ref.~\cite{Liu:2010xh}, in which is argued that the vector meson exchange 
plays an important role to create S wave bound states.  
On the other hand, in canonical quantization at mean field approximation
\cite{Freire:2010zd} the dynamic solution for the 
vector field is zero; the effective mass
of $D$-mesons is constant for low temperatures and 
almost zero only at very high temperature of $T$ = 1.2 GeV, in which the $D$ mesons should be dissociated. 

The usual deconfinement transition temperature for 
two flavors dynamical quarks is $T_c$ = 172 MeV \cite{Bornyakov:2011yb},
on the other hand lattice QCD in the quenched approximation is estimated 
the value of the deconfinement transition temperature which is 262.5 MeV \cite{Boyd:1996bx}.
In terms of the usual value of $T_c$ the suggested dissociation temperature in Ref. \cite{Freire:2010zd} is then $T$ = 1.2 GeV
$\approx 6.7 \, T_c$. Calculations in
lattice QCD, however, suggest that temperature of 
charmonium dissociation is a 
value between 1.6 $T_c$ and 2.35 $T_c$ \cite{Asakawa:2003re,Ding:2012sp}. 
In addition, a recent QCD sum rule result \cite{Gubler:2011ua}
predicts at temperature between 1.52 $T_c$ and 1.68 $T_c$
the charmonium should be dissociate. 

The effective mass of the hadron is the mass in which it 
seems to carry in the medium. If the effective mass is 
lower than rest mass, the hadron 
is found as bound state among it and the medium.
The interaction of nucleons in nuclear matter have
been studied using the Walecka model \cite{Walecka:1974qa}.
At finite temperature the interaction of many body 
nucleon-anti-nucleon system presents the effective mass 
of the nucleon is almost zero at 1.74 $T_c$ 
and a sea of pairs of N-$\bar{N}$ are produced in the
nuclear medium and this system behaves as massless bosons
system \cite{Walecka:2004}.
Another many-body cases are studied in
the interaction of mesons with nucleus,
such as bound state among meson and nucleus are 
called mesic nuclei has been explored in different ways:
D-nucleus \cite{Tsushima:1998ru,Yasui:2012rw,Kumar:2009xc}
$J/\psi$-nucleus \cite{Morita:2010pd}
$\rho$-nucleus \cite{Ghosh:2000zj},
K-nucleus \cite{Gazda:2012zz}. In these situations, the effective mass also appears as a relevant parameter to study the bound states, and has been used in Refs. \cite{Tsushima:1998ru,Kumar:2009xc,Morita:2010pd,Ghosh:2000zj}, 
where if effective meson mass decreases in the nucleus 
so mesic nuclei exist.

Hence, taking as motivation the analysis of exotic states within hadronic molecular approach and meson exchange models discussed above, in this work we apply the relativistic mean field theory 
developed in Refs. \cite{Walecka:1974qa,Walecka:2004,Freire:2010zd} 
to study the thermodynamic behavior of a system constituted of heavy-flavored mesons
in the framework of path-integral formalism. We use the following idea: 
whether the interaction among mesons builds a simple 
structure likes a di-hadron molecule, as suggested by Refs. \cite{Gamermann:2006nm,Ke:2012gm,Liu:2010xh,Li:2012mq,Zhang:2009vs}, 
so a many-body system of hadrons at high temperatures can form more complicate structures like a multi-meson system. In special, we analyze the effective mass of the $D$ meson, $m_{eff}$,
where if $m_{eff}$ decreases we have a bound of D in D mesons
matter or should have some bound states of $D-\bar{D}$ or X(3700)
in meson medium. In addition, if $m_{eff}$ goes to zero, 
a sea of pairs of $D\bar{D}$ or a multi-meson 
strongly interacting system is formed.

The organization of this paper is as follow. In Sec.~II, we introduce the Lagrangian density that describes a system of $D$-mesons and calculate the relevant quantities to study its  thermodynamic behavior. The analysis of thermodynamics of charmed meson matter is discussed in Sec.~III, while in Sec.~IV this approach is extended to bottomed meson matter. Finally, Section~V presents some concluding remarks.

\section{The Formalism}

We start by introducing the effective Lagrangian density that describes a system of $D$-mesons interacting with scalar ($\sigma$) and vector ($\omega$) mesons \cite{Ding:2008gr}, 
\ben
{\cal L} & = & (\partial _{\mu} D)(\partial ^{\mu} D^{\dag}) - m_{D}^2 D D^{\dag} - \frac{1}{4} F_{\mu \nu} F^{\mu \nu}+\frac{1}{2}m_{\omega}^2 \omega_{\mu} \omega^{\mu} \nonumber \\
& & + \frac{1}{2}(\partial _{\mu} \sigma)(\partial ^{\mu} \sigma) - \frac{1}{2}m_{\sigma}^2 \sigma^{2} + g_{D \bar{D}\sigma} D D^{\dag} \sigma + i g_{D \bar{D}\omega} \omega^{\mu} [ D\partial_{\mu} D^{\dag} - (\partial_{\mu} D)D^{\dag} ], 
\label{Lag}
\een
where $F^{\mu \nu} = \partial ^{\mu} \omega^{\nu} - \partial ^{\nu} \omega^{\mu}$, $D = (D^0,D^+) $ is the doublet of the $D$ meson field, $m_{D}, m_{\omega}$ and $m_{\sigma}$ are the respective masses of $D$, $\sigma$ and $\omega$ mesons, and $g_{D \bar{D}\sigma}$ and $g_{D \bar{D}\omega}$ are the coupling constants for the $DD\sigma$ and $DD\omega$ interactions. 

In Lorentz gauge,  i.e. $\partial _{\mu} \omega^{\mu} = 0$, the equations of motion obtained from Eq. (\ref{Lag}) are
\ben
\partial _{\mu}\partial ^{\mu} \sigma + m_{\sigma}^2 \sigma & = & g_{D \bar{D}\sigma}  D D^{\dag} \label{Eq_Motiont1}, \\
\partial _{\mu}\partial ^{\mu} \omega^{\nu} + m_{\sigma}^2 \omega^{\nu} & = & -i g_{D \bar{D}\omega}  [ D\partial ^{\nu} D^{\dag} - (\partial ^{\nu} D)D^{\dag} ] \label{Eq_Motiont2}, \\
\partial^{\prime} _{\mu}\partial ^{ \prime \mu} D + (m_{D}^2 - g_{D \bar{D}\sigma}  \sigma + g_{D \bar{D}\omega}^2 \omega ^{\mu} \omega _{\mu}) D & = & 0, \label{Eq_Motiont3}
\een
where $\partial^{\prime} _{\mu} = \partial _{\mu} + i g_{D \bar{D}\omega} \omega_{\mu}$.

For our purposes, it is convenient to work in the context of the mean-field approximation \cite{Walecka:1974qa,Freire:2010zd}, replacing $\sigma$ and $\omega$ meson fields by classical fields,
\ben
\bar{\sigma} = \left\langle \sigma \right\rangle, \nonumber \\
\bar{\omega} = \left\langle \omega ^0 \right\rangle,
\label{approx}
\een
with $\omega ^{\mu} = 0$ for $\mu \neq 0$. This approximation allows us to rewrite the Lagrangian density in Eq. (\ref{Lag}) as
\ben
{\cal L} & = & D \left[ -\partial _{\mu}\partial ^{\mu} - (m_{D}^2 - g_{D \bar{D}\sigma} \sigma ) \right] D^{\dag} +\frac{1}{2}m_{\omega}^2 \bar{\omega}^{2} - \frac{1}{2}m_{\sigma}^2 \bar{\sigma}^{2} \nonumber \\
& & - g_{D \bar{D}\omega} \; \bar{\omega} \; j^0,
\label{Lag2}
\een
where $j^0 = i[ D\partial ^{0} D^{\dag} - (\partial ^{0} D)D^{\dag} ] $. Besides, it can be seen from Eqs. (\ref{Eq_Motiont1})-(\ref{Eq_Motiont3}) in mean-field approximation that $\bar{\sigma} \propto \rho_s $ and $\bar{\omega} \propto \rho $, with $\rho _s $ and $\rho$ being the scalar and number densities, respectively. 
 
To study the thermodynamic behavior of the system introduced above, we assume that it is in thermodynamic equilibrium and at finite temperature $T$ and density $\mu$ of $D$-mesons. In this sense, we use imaginary time (Matsubara) formalism, and the grand partition function takes the form 
\cite{EE},
\be
{\cal Z} \propto \int {\cal D} D \,\,{\cal D} D^{\dag} \exp{\left\{ -\int _{0}^{\beta} d\tau \int d^3 x \left[ {\cal L}_{E} + \mu j_0 \right]\right\}} , 
\label{PartFunct}
\ee
where $\beta = 1/T $ and $\mu$ are the inverse of temperature and chemical potential, respectively, and ${\cal L}_{E}$ is the Lagrangian density given by Eq. (\ref{Lag2}) in Euclidean space \footnote{Although the omission of the steps of derivation of Eq. (\ref{PartFunct}), it is worth to notice that this system is non-singular and there are no constraints, since the velocities obtained from conjugate momenta are primarily expressible (the usual constraint, the Lorentz gauge, is not relevant in the mean-field approximation we have adopted). Thus, the Hamiltonian density is constructed in an usual way, and after the integration over the momenta the grand partition function assumes the form in Eq. (\ref{PartFunct}).}.
Then, after the integration over the fields $ D $ and $ D ^{\dag }$ the obtained thermodynamic potential becomes,
\ben
\Omega (T, \mu) & = & V \left[ \frac{1}{2}m_{\sigma}^2 \bar{\sigma}^{2} -\frac{1}{2}m_{\omega}^2 \bar{\omega}^{2} \right] \nonumber \\
& & + \frac{V}{\beta}  \sum _{n = - \infty}^{\infty}\int \frac{d^3 p}{(2 \pi)^3} \ln{\left\{ \left[ \frac{2 \pi  n}{\beta } - i \mu _{eff} \right]^2 + p^2 + m_{eff}^2 \right\} },
\label{pot1}
\een
where the effective mass and effective potential of the $D$ mesons in hadronic medium read, respectively,
\ben 
m_{eff}^2 & = & m_{D}^2 - g_{D \bar{D}\sigma}  \bar{\sigma} ,\nonumber \\
\mu _{eff} & = & \mu - g_{D \bar{D}\omega}\bar{\omega} .
\label{eff}
\een
$V$ is the volume. 

In order to obtain the thermodynamic potential in a more tractable form, we use zeta-function regularization techniques
\cite{EE,Inagaki:1994ec,Abreu:2009zz,Abreu:2011rj}. 
In this scenario, Eq. (\ref{pot1}) can be rewritten as 
\ben
\Omega (T, \mu) & = & V \left( \frac{1}{2}m_{\sigma}^2 \bar{\sigma}^{2} -\frac{1}{2}m_{\omega}^2 \bar{\omega}^{2} \right) + \frac{V}{\beta} \zeta ^{\; \prime} (0),
\label{pot1a}
\een
where the zeta function $\zeta (s)$ is given by 
\ben
\zeta (s) & = &  \sum_{n =  -\infty}^{+\infty }
\int \frac{ d ^3 p }{(2\pi)^{3}}\left[ \left( \frac{2 \pi  n}{\beta } - i \mu _{eff} \right)^2 + p^2 + m_{eff}^2  \right]^{-s } \nonumber \\
& = &  \frac{\Gamma \left( s- \frac{3}{2} \right)}{(4\pi)^{3/2} \Gamma \left( s \right) }  \sum_{n =  -\infty}^{+\infty } \left[ \left( \frac{2 \pi  n}{\beta } - i \mu _{eff} \right)^2 + m_{eff}^2  \right]^{-s + \frac{3}{2} }.  
\lb{zeta1}
\een
Notice that $\zeta ^{\; \prime} (s)$ denotes the derivative of $\zeta$ with respect to the argument $s$. 

Then, by using the identity 
\begin{eqnarray}
\sum_{n =  -\infty}^{+\infty } \left[ a \left( n - c \right)^2 + q^2  \right]^{-\eta} & =&   \sqrt{\frac{\pi}{a}} \;\frac{ \Gamma \left(\eta -\frac{1}{2}\right)}{ \Gamma (\eta)} q^{1-2\eta }
\nonumber \\
& &  + \frac{4 \pi^{\eta }}{ \sqrt{a} \;\Gamma \left(\eta\right)} \sum_{n =  1}^{\infty }\cos{(2\pi n c)} \left( \frac{n}{\sqrt{a} \; q} \right)^{\eta - \frac{1}{2}}
K_{\eta - \frac{1}{2}} \left( \frac{2 \pi n q}{\sqrt{a}} \right) , 
\label{epstein2}
\end{eqnarray}
and after some manipulations the thermodynamic potential in Eq. (\ref{pot1a}) becomes
\begin{eqnarray}
\Omega(T, \mu) & = & V \left( \frac{1 }{2} m_{\sigma} \bar{\sigma} ^2 - \frac{1 }{2} m_{\omega}\bar{\omega}^2 \right) + U _{vac}  \nonumber \\
& & - \frac{V}{\pi ^{2} }\sum _{n = 1}^{\infty} \cosh{( \beta n \mu _{eff} ) } \left( \frac{m_{eff}}{n \beta} \right)^2 K_{2}(n \beta m_{eff}),
\label{pot2}
\end{eqnarray}
where $K_{\nu}$ is the modified Bessel function of second kind, and $U _{vac} $ is the $T,\mu$-independent vacuum contribution which comes from the pole term in Eq. (\ref{epstein2}), 
\begin{eqnarray}
U _{vac} = \frac{1}{64 \pi ^2} \left\{2 m_{eff}^2 \Lambda ^2-\Lambda ^4+2 m_{eff}^4 \ln{\left(m_{eff}^2\right)}+2 \left(\Lambda ^4 - m_{eff}^4\right) \ln{\left(m_{eff}^2+\Lambda ^2\right)}\right\}.
\label{pot_vac}
\end{eqnarray}
with $\Lambda $ being a cutoff. We will omit this term henceforth.

Another requirement necessary to the thermodynamics of this model is the analysis of gap equations,
\begin{eqnarray}
\frac{ \partial \Omega }{ \partial \bar{\sigma}} & = & 0, \label{gap1} \\
 \frac{ \partial \Omega }{ \partial \bar{\omega} } & = & 0. \label{gap2}
\end{eqnarray}
These expressions give the values for $\bar{\sigma}$ and $\bar{\omega}$ that extremize the thermodynamic potential $\Omega$. So, the use of Eq. (\ref{pot2}) in gap equations (\ref{gap1}) and (\ref{gap2}) yields 
\begin{eqnarray}
\bar{\sigma} & = & \frac{ g_{D \bar{D}\sigma} }{ m_{\sigma}^{2} } \rho_s,
\label{gap3} \\
 \bar{\omega} & =&  - \frac{ g_{D \bar{D}\omega} }{ m_{\omega}^{2} }  \rho , 
\label{gap4}
\end{eqnarray}
where the scalar and number densities are written as 
\begin{eqnarray}
\rho_s & = & \frac{ m_{eff}}{2 \pi^{2} \beta}\sum _{n = 1}^{\infty} \cosh{(\beta n \mu _{eff} )} \frac{1}{n} K_{1}(n \beta m_{eff}),
\label{dens1} \\
\rho & =&  - \frac{ m_{eff}^2}{\pi^{2} \beta}\sum _{n = 1}^{\infty} \sinh{(\beta n \mu _{eff} )} \frac{1}{n} K_{2}(n \beta m_{eff})
\label{dens2}.
\end{eqnarray}
As remarked before, we have omitted the contribution coming from vacuum term.

It is worthy mentioning that if the numbers of $D$ and $\bar{D}$ mesons are equal in hadronic matter, which means a vanishing effective chemical potential $\mu_{eff} = 0$, from Eqs. (\ref{gap4}) and (\ref{dens1}) we see that $\rho = 0$ and therefore $ \bar{\omega} = 0$, just as the situation reported in 
Ref. \cite{Freire:2010zd}. 

Now we derive other relevant thermodynamic quantities from the thermodynamic potential given by Eq. (\ref{pot2}). For example, the pressure reads
\begin{eqnarray}
p(T, \mu) & \equiv & - \frac{\partial \Omega}{\partial V} \nonumber \\
& = &  \frac{1 }{2} m_{\omega}\bar{\omega}^2 - \frac{1 }{2} m_{\sigma} \bar{\sigma} ^2 - U _{vac}  \nonumber \\
& & + \frac{1}{\pi ^{2} }\sum _{n = 1}^{\infty} \cosh{( \beta n \mu _{eff} ) } \left( \frac{m_{eff}}{n \beta} \right)^2 K_{2}(n \beta m_{eff}),
\label{pressure}
\end{eqnarray}
Therefore, the values of $\bar{\sigma}$ and $\bar{\omega}$ that extremize the pressure must also satisfy Eqs. (\ref{gap3}) and (\ref{gap4}).

In addition, the entropy and energy densities at chemical equilibrium  are given by,   
\begin{eqnarray}
s(T) & \equiv & \left.\frac{\partial p}{\partial T} \right|_{\mu _{eff} = 0} \nonumber \\
& = &  \frac{1}{\pi ^{2} }\sum _{n = 1}^{\infty} \frac{m_{eff}^3}{n }  K_{3}(n \beta m_{eff})  ,
\label{entropy}
\end{eqnarray}
and
\begin{eqnarray}
\varepsilon(T) & \equiv & \left( - p + T s \right)_{\mu _{eff} = 0} \nonumber \\
& = &  \frac{1 }{2} m_{\sigma} \bar{\sigma} ^2   \nonumber \\
& & + \frac{1}{\pi ^{2} }\sum _{n = 1}^{\infty} \left[ 3 \left( \frac{m_{eff}}{n \beta} \right)^2 K_{2}(n \beta m_{eff}) + \frac{m_{eff}^3}{n \beta}  K_{1}(n \beta m_{eff})  \right],
\label{energy}
\end{eqnarray}
respectively.

\section{Effective mass of $D$-mesons}

We now analyze the thermodynamic behavior of the model introduced in previous section under the change of values of the relevant parameters. We consider the situation $\mu_{eff} = 0$, where the number of $D$ and $\bar{D}$ mesons are equal in mesonic matter. We use the following parameters: $m_D = 1.87$ GeV, $m_{\sigma} = 0.5$ GeV \cite{Beringer:1900zz}, $g_{D \bar{D}\sigma} = 2.85$ GeV \cite{Ding:2008gr} and deconfinement 
temperature $T_c$ = 172 MeV  \cite{Bornyakov:2011yb}. 

\begin{figure}[!h]
\includegraphics[{height=7.0cm}]{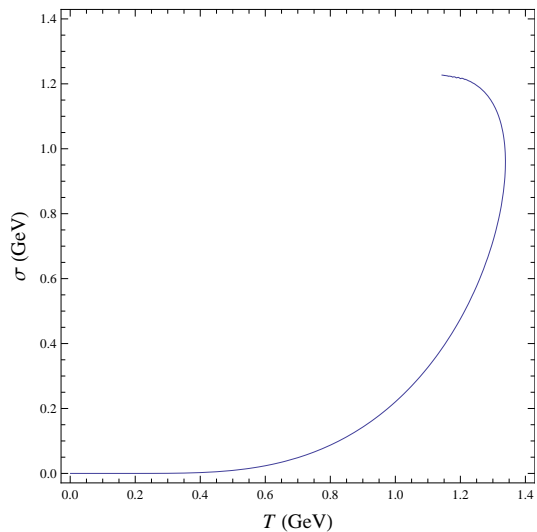}
\caption{Plot of $\bar{\sigma}$ in Eq.~(\ref{gap3}), as a function of temperature, at chemical equilibrium. 
} 
\label{fig0}
\end{figure}

In Fig.~\ref{fig0} is plotted the sigma field in Eq. (\ref{gap3}) as a function of temperature. We note that $\bar{\sigma}$ does not exist for temperatures above of $T = 1.34$ GeV. 

Besides,  $\bar{\sigma}$ has a maximum value at $T \equiv T_D \approx 1.15$ GeV or $T_D \approx 6.68 T_c$, which is the critical value in which we have an almost vanishing effective mass. 
This critical temperature can be obtained analytically by taking $m_{eff} \approx 0$ in Eq. (\ref{gap3}) and using the asymptotic formula of Bessel functions, that is
\be
T_D = \sqrt{12}\,\frac{m_D\,m_\sigma}{g_{D \bar{D}\sigma}}.
\label{Tcri}
\ee 
Replacing the values of the parameters above mentioned, we get $T_D \approx 1.137$ GeV or 6.6 $T_c$, in agreement with the Fig.~\ref{fig0}.

\begin{figure}[!h]
\includegraphics[{height=7.0cm}]{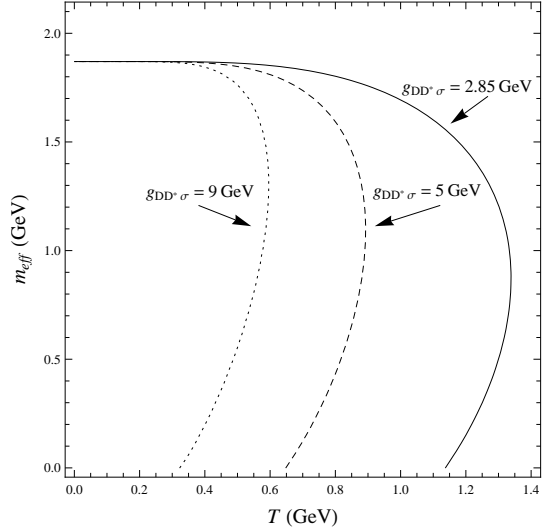}
\caption{Plot of effective mass of $D$ mesons in Eq.~(\ref{gap3}) as a function of temperature, at chemical equilibrium, for different values of coupling constant. 
} 
\label{fig10}
\end{figure}

Let us now turn our attention to the effective mass of the $D$ mesons in hadronic matter. We plot in Fig.~\ref{fig10} the values for $m_{eff}$ as a function of temperature which we solve the gap equation (\ref{gap3}). As in Ref. \cite{Freire:2010zd}, we can see that the effective mass reduces as the temperature increases, and a transition from interacting gas to very strongly interacting matter appears at $T_D$. 
Moreover, this temperature is too high compared 
with the temperature of charmonium dissociation, where its temperature is a value between 1.6 $T_c$ and 2.35 $T_c$ \cite{Asakawa:2003re,Ding:2012sp}. 
So, in this scenairo the $D$-mesons should be dissociated. 

We can estimate binding energy of a meson molecule composed of two mesons $D$ and $\bar{D}$, defined by 
\be
\epsilon (T) = m_{D\bar{D}} -2 \,m_{eff}(T),
\label{bind_energy}
\ee
where $m_{D\bar{D}}\equiv 2\,m_D$ and $2 \,m_{eff}(T)$ is the mass of $D\bar{D}$-bound state. Then, considering the maximum temperature of 2.09 $T_c$ in Fig.~\ref{fig10}, the effective mass has a value 1.869 GeV. Therefore, a $D\bar{D}$-meson molecule has a bind energy of 2 MeV. For the sake of comparison, notice that in Ref.~\cite{Liu:2010xh} is reported the binding energy of 1.4 MeV for the S wave $D\bar{D}$ bound state at $T=0$ (with an cutoff $\Lambda$= 1.5 GeV and $m_{\sigma}$= 400 MeV). Thus, this result is close with our findings, but in our formalism the bound 
state only could be appear at $T>0$. Note also that if we consider $m_{\sigma}$= 400 MeV at temperature 1.92 $T_c$ we get the same effective mass of 1.869 GeV.

In Fig. (\ref{fig10}) is plotted the effective mass as function of temperature by taking bigger values of the coupling constant $g_{D \bar{D}\sigma}$ with respect to the vacuum value of $g_{D \bar{D}\sigma}$. 
It can be remarked that the critical temperature decreases as $g_{D \bar{D}\sigma}$ increases, as explicited in Eq. (\ref{Tcri}). For instance, with $g_{D\bar{D}\sigma} =$ 9 GeV in Eq. (\ref{Tcri}) we get $T_{D} \approx 0.36$ GeV or 2.09 $T_c$, 
in agreement with Fig.(\ref{fig10}). 
This fact explicits the role of the sigma field in binding the $D$ mesons, and the increasing of the magnitude of interaction among the sigma and $D$-meson fields forces the system to experiment a phase transition at smaller temperatures.

\begin{figure}[!h]
\includegraphics[{height=7.0cm}]{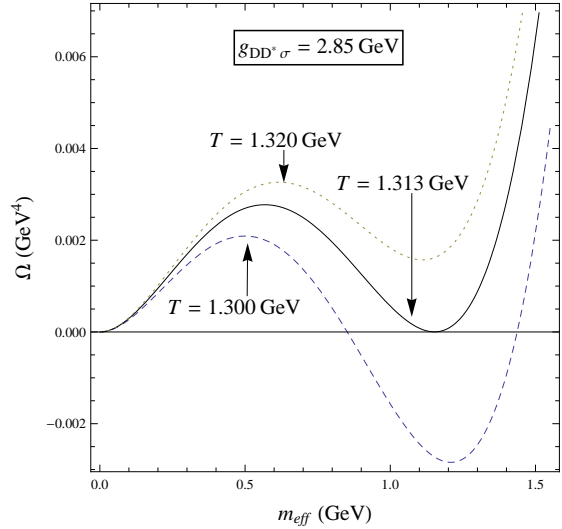}
\caption{Plot of thermodynamic potential of $D$ mesons in Eq.~(\ref{pot2}) as a function of temperature, at chemical equilibrium and for $g_{D \bar{D}\sigma}$ = 2.85 GeV. Dashed, solid and dotted lines represent the cases for $T = 1.300$, 1.313 and 1.320 GeV, respectively. } 
\label{fig2}
\end{figure}

\begin{figure}[!h]
\includegraphics[{height=7.0cm}]{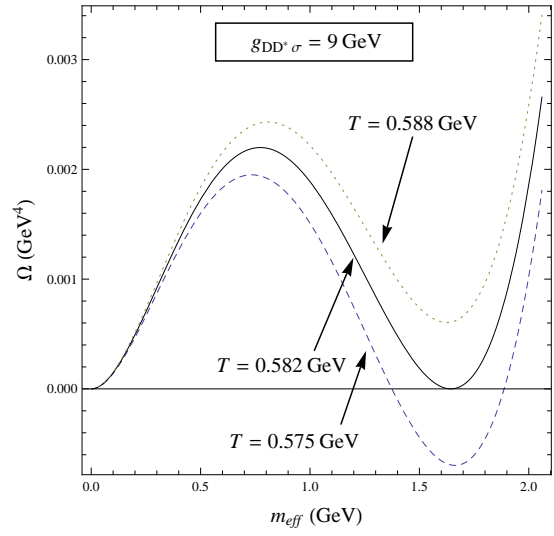}
\caption{Plot of thermodynamic potential of $D$ mesons in Eq.~(\ref{pot2}) as a function of temperature, at chemical equilibrium and for $g_{D \bar{D}\sigma}$ = 9 GeV. Dashed, solid and dotted lines represent the cases for $T = 0.575$, 0.582 and 0.588 GeV, respectively. } 
\label{fig2b}
\end{figure}

For a better understanding of the behavior of this system, in Figs.~\ref{fig2} and ~\ref{fig2b} are plotted the thermodynamic potential written in Eq. (\ref{pot2}) as function of effective mass for $g_{D \bar{D}\sigma}$ = 2.85 GeV and 9 GeV, respectively. As suggested by the previous figures, we observe a first-order phase transition occurring at $T_0 \approx 1.31$ GeV for $g_{D \bar{D}\sigma}$ = 2.85 GeV, and $T_0 \approx 0.582$ GeV for $g_{D \bar{D}\sigma}$ = 9 GeV. 

\begin{figure}[!h]
\includegraphics[{height=7.0cm}]{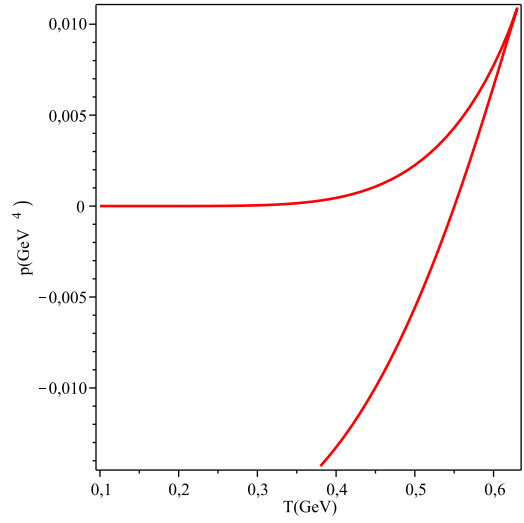}
\caption{Pressure of $D$-meson matter as a function of temperature
at  $g_{D \bar{D}\sigma}$ = 9 GeV. } 
\label{pressure_g9}
\end{figure}

\begin{figure}[!h]
\includegraphics[{height=7.0cm}]{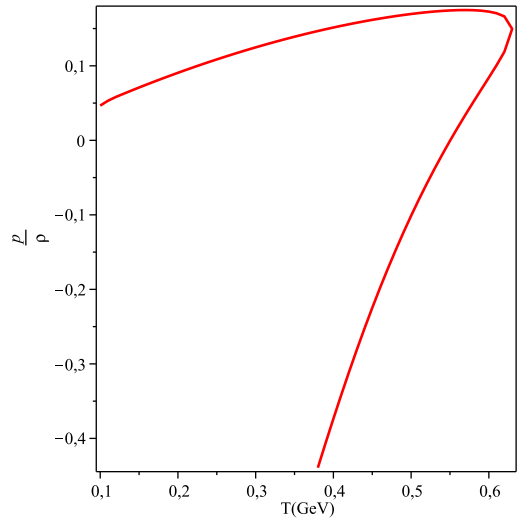}
\caption{Ratio of pressure/energy density of $D$-meson matter as a function of temperature
at  $g_{D \bar{D}\sigma}$ = 9 GeV.} 
\label{ratio_g9}
\end{figure}

In Fig.~\ref{pressure_g9} is plotted the pressure in Eq.~(\ref{gap3}) as a function of temperature for 
$g_{D \bar{D}\sigma}$ = 9 GeV. We note an exotic behavior for values $T>$ 2.09 $T_c$, 
in which the pressure will assume negative values. On the other hand, as remarked in Ref. \cite{Freire:2010zd}, for $g_{D \bar{D}\sigma}$ = 2.85 GeV the pressure has only positive values. 
Thus, this result suggests that a greater interaction between the $D$-mesons
induces a phase transition from gas into
a highly interacting matter with negative pressure.
Another interesting fact is that the ratio of pressure
to energy density given in Fig.~\ref{ratio_g9}
is in a range of values typical to that predicted for matter that has a behavior similar to dark energy, i.e.
$-1<\frac{p}{\rho}<-1/3$ \cite{Padmanabhan:2004av,Albrecht:2006um}.

\begin{figure}[!h]
\includegraphics[{height=7.0cm}]{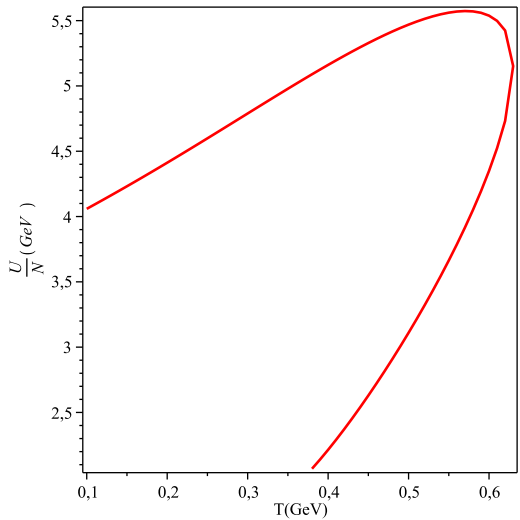}
\caption{Energy per pair of $D$-meson matter as a function of temperature
at  $g_{D \bar{D}\sigma}$ = 9 GeV.} 
\label{U_N_g9}
\end{figure}

In addition, another exotic feature of the system with
$g_{D \bar{D}\sigma}$ = 9 GeV
is that the energy per pair of mesons at temperature $T_ {D}$ 
has a phase with an energy of 3 GeV less than the gas 
phase, as can be seen from Fig.~\ref{U_N_g9}.

\newpage
\section{Effective mass of $B$-mesons}

Taking as a guiding principle the heavy quark symmetry, a natural question that appears is about the extension of the approach discussed in previous Sections to systems constituted of other heavy-flavored mesons. Then, in the following we study the bottomonium analogously to the charmonium situation analyzed above. To do so, we just replace the $D$-mesons with the corresponding $B$-mesons in the model displayed in Eq.(\ref{Lag}): the doublet of $B$-mesons is $B = (B^-,\bar{B}^0) $. In the present case we use the following parameters: $m_B = 5.279$ GeV \cite{Beringer:1900zz} and 
$g_{B \bar{B}\sigma} = 8.04$ GeV \cite{Ding:2008gr}.

Thus, using Eq.~(\ref{Tcri}) adapted to $B$-meson system, we get 
$T_{B} \approx 1.137$ GeV or $6.6\, T_c$, i.e. equal to the temperature of the $D$-mesons $T_D$.  It deserves mention that this is so because the coupling constants $g_{D \bar{D}\sigma}$ and $g_{B \bar{B}\sigma} $ we have used are proportional to the mass of the respective heavy meson, as shown in Ref. \cite{Ding:2008gr}, which leave the temperatures $T_D$ and $T_B$ independent of the mass of heavy-flavored meson. 

Furthermore, we note that the temperature $T_B$ is much higher than the values
suggested in literature for the dissociation of bottomonium, which are:
2.06 $T_c$ \cite{Jakovac:2006sf} and  4.18 $T_c$ \cite{Emerick:2011xu,Rapp:2008tf}.

\begin{figure}[!h]
\includegraphics[{height=7.0cm}]{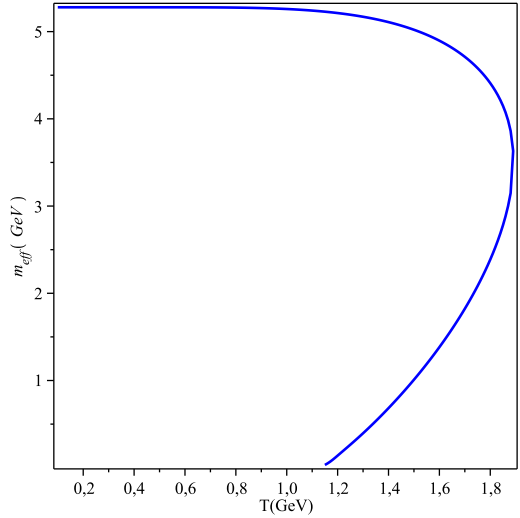}
\caption{Plot of effective mass of $B$-meson as a function of temperature, at chemical equilibrium, for different values of coupling constant. 
} 
\label{massB}
\end{figure}

In Fig.~\ref{massB} is plotted the effective mass as a function of temperature.
We observe the effective mass of B meson goes to 
zero at temperatures close to the value $T_{B}$.
If we consider the maximum temperature given by 3.953 $T_c$ in Fig.~\ref{massB} the effective mass has a value of 5.278 GeV. 
Here, a meson molecule composed of two mesons $B$ and $\bar{B}$ 
has a binding energy given by $2(m_{B}-\,m_{eff})$ or 2 MeV, which is equal to that obtained for the $D\bar{D}$-molecule. On the other hand, it does not agree to the results of Ref.~\cite{Liu:2010xh} which provide a value of 60.7 MeV for the bound state $B\bar{B}$. In our case to obtain a value of
60 MeV for this bound state, we must consider that 
the molecule survives at temperature 6.16 $T_c$, which is slightly 
lower than the critical temperature $T_ {B}$.

\begin{figure}[!h]
\includegraphics[{height=7.0cm}]{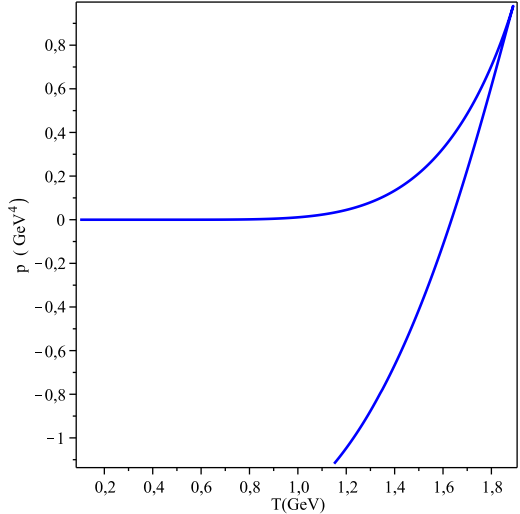}
\caption{Pressure of the system of $B$-mesons as a function of temperature
at  $g_{B \bar{B}\sigma}$ = 8.04 GeV.} 
\label{pressureB}
\end{figure}

In Fig.~\ref{pressureB} is plotted the pressure as a function of temperature. 
We note that the pressure has the negative values for $T>T_{B}$,
even working with the coupling obtained in vacuum.
This result suggests that the interaction among the $B$-mesons at temperature
$T>T_{B}$ is stronger than the situation at zero-temperature.

\begin{figure}[!h]
\includegraphics[{height=7.0cm}]{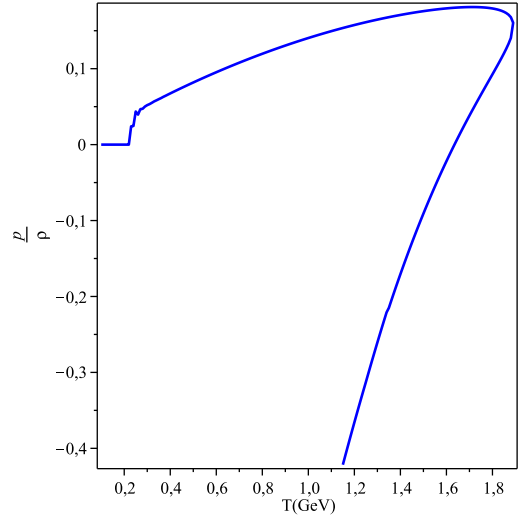}
\caption{Ratio of pressure/energy density of $B$-mesons as a function of temperature
at $g_{B \bar{B}\sigma}$ = 8.04 GeV.} 
\label{ratioB}
\end{figure}

The ratio of pressure to energy density is shown in Fig.~\ref{ratioB}. Differently to the case of $D$-mesons, with the vacuum coupling constant ($g_{B \bar{B}\sigma}$ = 8.04 GeV) this ratio present values in the range predicted for matter that behaves as dark energy one.

\begin{figure}[!h]
\includegraphics[{height=7.0cm}]{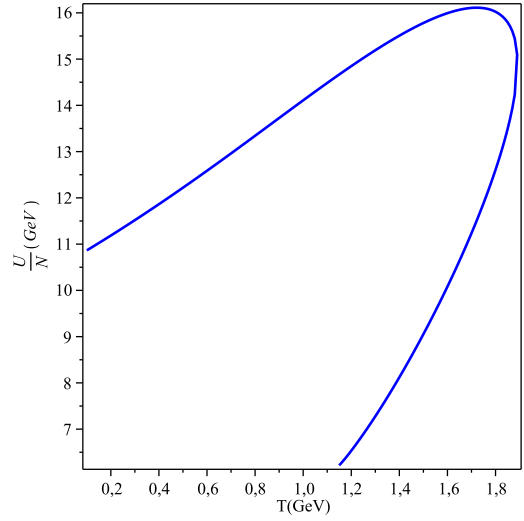}
\caption{Energy per pair of $B$-mesons as a function of temperature
at  $g_{B \bar{B}\sigma}$ = 8.04 GeV } 
\label{energyB}
\end{figure}

Finally, another exotic feature can be pointed from Fig.~\ref{energyB}, in which is plotted the energy per pair of mesons versus temperature at vacuum coupling constant. We see that the energy per pair of mesons at temperature $T_{B}$ has a phase with an energy of 8 GeV less than the gas phase.

\newpage
\section{Concluding Remarks}

In this work the thermodynamic behavior of a system constituted of heavy-flavored mesons has been evaluated, with the $D\bar{D}$ or $B\bar{B}$ interaction being studied within a $\sigma,\omega$-meson exchange model. At mean-field approximation, 
we have shown that the omega meson does not contribute when 
the system has the same number of heavy mesons and anti-mesons. On the other hand, 
the sigma field has nontrivial solution coming from gap equation, and grows as the temperature increases.

Using the value of coupling constant typical of the interaction between $D$ and $\bar{D}$
in vacuum, $g_{D\bar{D}\sigma} =2.85 $ GeV, our results suggest that the effective mass
decreases as temperature is raised. Also, a bound state with energy of 2 MeV can be formed at temperature of 2.09 times the deconfinement temperature $T_c$, which is slightly above the values predicted in literature for the charmonium dissociation.

In addition, an interesting result has been obtained for bigger values of vacuum coupling constant,
where above the temperature of 2.09 $T_c$, we have obtained a strongly interacting phase having a negative pressure, and a ratio of pressure to energy density in the range of values typical of systems that present behaviors like dark energy matter.  
Notice, however, that since the coupling constant $g_{D\bar{D}\sigma}$ is not free, we believe that the charmed mesonic matter should not show negative pressure. Nevertheless, this result might be useful in models of dark energy in which interacting fields could be treated in usual way, 
as an alternative for instance to the noncommutative field theory \cite{Barosi:2008gx}.

We have also analyzed bottomed mesonic matter analogously to the charmed system. With the value of coupling constant $g_{B\bar{B}\sigma}$ in vacuum, at temperature of 3.953 $T_c$ the bound state 
$B\bar{B}$ has a binding energy of 2 MeV. 
This result is lower than the result of Ref.~\cite{Liu:2010xh},  
which provide a value of 60.7 MeV.
For a value of 60 MeV for the bound state in our case, 
we must consider a molecule survives at temperature 6.16 $T_c$, 
which is above the temperature of bottomonium dissociation 
available in literature: 4.18 $T_c$~\cite{Emerick:2011xu,Rapp:2008tf}.

For the case of matter of $B$-mesons at temperature up to 6.6 $T_c$,
we again obtain a phase with negative pressure and
ratio of pressure to energy density within the range of values
expected for systems with behavior like dark energy matter. It is relevant to notice, though, that the negative pressure can be discarded if the bottomonium dissolve at lower temperatures
than 6.6 $T_c$, as have been predicted by 
Refs.~\cite{Jakovac:2006sf,Emerick:2011xu,Rapp:2008tf}.


\section{Acknowledgements}
{We would like to thank Profs. Alexandre L. Gadelha, 
Francisco A. Brito and Tiago Mariz for fruitful discussions.
We are grateful to Dr. Philipp Gubler for the explanations 
of the value of $T_c$ used in lattice QCD and QCDSR.
This work has been partially supported by CAPES-PROCAD (Brazil).}
\vfill \eject


\begin{references}

\bibitem{Choi:2003ue} 
  S.~K.~Choi {\it et al.}  [Belle Collaboration],
  Phys.\ Rev.\ Lett.\  {\bf 91}, 262001 (2003)
  [hep-ex/0309032].
    
\bibitem{Abe:2004zs} 
  K.~Abe {\it et al.}  [Belle Collaboration],
  Phys.\ Rev.\ Lett.\  {\bf 94}, 182002 (2005)
  [hep-ex/0408126].

\bibitem{Choi:2007wga} 
  S.~K.~Choi {\it et al.}  [BELLE Collaboration],
  Phys.\ Rev.\ Lett.\  {\bf 100}, 142001 (2008)
  [arXiv:0708.1790 [hep-ex]].

\bibitem{Aaltonen:2009tz} 
  T.~Aaltonen {\it et al.}  [CDF Collaboration],
  Phys.\ Rev.\ Lett.\  {\bf 102}, 242002 (2009)
  [arXiv:0903.2229 [hep-ex]].

\bibitem{Aaij:2011sn} 
  R.~Aaij {\it et al.}  [LHCb Collaboration],
  Eur.\ Phys.\ J.\ C {\bf 72}, 1972 (2012)
  [arXiv:1112.5310 [hep-ex]].


\bibitem{Godfrey:1985xj} 
  S.~Godfrey and N.~Isgur,
  Phys.\ Rev.\ D {\bf 32}, 189 (1985).


\bibitem{Beringer:1900zz} 
  J.~Beringer {\it et al.}  [Particle Data Group Collaboration],
  Phys.\ Rev.\ D {\bf 86}, 010001 (2012).

\bibitem{Seth:2004zb} 
  K.~K.~Seth,
  Phys.\ Lett.\ B {\bf 612}, 1 (2005)  [hep-ph/0411122].  

\bibitem{Morningstar:1999rf} 
  C.~J.~Morningstar and M.~J.~Peardon,
Phys.\ Rev.\ D {\bf 60}, 034509 (1999)  [hep-lat/9901004].  

\bibitem{Barnes:2003vb} 
  T.~Barnes and S.~Godfrey,
  Phys.\ Rev.\ D {\bf 69}, 054008 (2004)  [hep-ph/0311162]. 

\bibitem{Maiani:2004vq} 
  L.~Maiani, F.~Piccinini, A.~D.~Polosa and V.~Riquer,
  Phys.\ Rev.\ D {\bf 71}, 014028 (2005)
  [hep-ph/0412098].

\bibitem{Tornqvist:1993ng} 
  N.~A.~Tornqvist,
  Z.\ Phys.\ C {\bf 61}, 525 (1994)
  [hep-ph/9310247].

\bibitem{Voloshin:1976ap} 
  M.~B.~Voloshin and L.~B.~Okun,
  JETP Lett.\  {\bf 23}, 333 (1976)
  [Pisma Zh.\ Eksp.\ Teor.\ Fiz.\  {\bf 23}, 369 (1976)].

\bibitem{Novikov:1977dq} 
  V.~A.~Novikov, L.~B.~Okun, M.~A.~Shifman, A.~I.~Vainshtein, M.~B.~Voloshin and V.~I.~Zakharov,
  Phys.\ Rept.\  {\bf 41}, 1 (1978).

\bibitem{Gamermann:2006nm} 
  D.~Gamermann, E.~Oset, D.~Strottman and M.~J.~Vicente Vacas,
  Phys.\ Rev.\ D {\bf 76}, 074016 (2007)
  [hep-ph/0612179].

\bibitem{Ablikim:2006zq} 
  M.~Ablikim {\it et al.}  [BES Collaboration],
Phys.\ Rev.\ Lett.\  {\bf 97}, 121801 (2006)  [hep-ex/0605107].  

\bibitem{Ma:2008nj}
  H.~L.~Ma [BES Collaboration],
  arXiv:0810.3541 [hep-ex]. 

\bibitem{Rong:2010ai} 
  G.~Rong, D.~Zhang and J.~C.~Chen,
  arXiv:1003.3523 [hep-ex].
  
\bibitem{Voloshin:2005sd} 
  M.~B.~Voloshin,
  Phys.\ Rev.\ D {\bf 71}, 114003 (2005)
  [hep-ph/0504197].
  
\bibitem{Ke:2012gm} 
  H.~-W.~Ke, X.~-Q.~Li, Y.~-L.~Shi, G.~-L.~Wang and X.~-H.~Yuan,
  JHEP {\bf 1204}, 056 (2012)
  [arXiv:1202.2178 [hep-ph]].

\bibitem{Liu:2010xh} 
  Y.~-R.~Liu, M.~Oka, M.~Takizawa, X.~Liu, W.~-Z.~Deng and S.~-L.~Zhu,
  Phys.\ Rev.\ D {\bf 82}, 014011 (2010)
  [arXiv:1005.2262 [hep-ph]].
 
\bibitem{Li:2012mq} 
  M.~T.~Li, W.~L.~Wang, Y.~B.~Dong and Z.~Y.~Zhang,
  Int.\ J.\ Mod.\ Phys.\ A {\bf 27}, 1250161 (2012)
  [arXiv:1206.0523 [nucl-th]].

\bibitem{Zhang:2009vs} 
  J.~-R.~Zhang and M.~-Q.~Huang,
  Phys.\ Rev.\ D {\bf 80}, 056004 (2009)
  [arXiv:0906.0090 [hep-ph]].

\bibitem{Freire:2010zd} 
  M.~L.~d.~F.~Freire and R.~R.~da Silva,
  AIP Conf.\ Proc.\  {\bf 1296}, 346 (2010)
  [arXiv:1003.1690 [hep-ph]].

\bibitem{Yang:2011rp} 
  Y.~Yang, J.~Ping, C.~Deng and H.~-S.~Zong,
  J.\ Phys.\ G {\bf 39}, 105001 (2012)
  [arXiv:1105.5935 [hep-ph]].

\bibitem{Bornyakov:2011yb}
  V.~G.~Bornyakov, R.~Horsley, Y.~Nakamura, M.~I.~Polikarpov, P.~Rakow and G.~Schierholz,
  arXiv:1102.4461 [hep-lat].

\bibitem{Boyd:1996bx} 
  G.~Boyd, J.~Engels, F.~Karsch, E.~Laermann, C.~Legeland, M.~Lutgemeier and B.~Petersson,
  Nucl.\ Phys.\ B {\bf 469}, 419 (1996)
  [hep-lat/9602007].

\bibitem{Asakawa:2003re} 
  M.~Asakawa and T.~Hatsuda,
  Phys.\ Rev.\ Lett.\  {\bf 92}, 012001 (2004)
  [hep-lat/0308034].

\bibitem{Ding:2012sp} 
  H.~T.~Ding, A.~Francis, O.~Kaczmarek, F.~Karsch, H.~Satz and W.~Soeldner,
  Phys.\ Rev.\ D {\bf 86}, 014509 (2012)
  [arXiv:1204.4945 [hep-lat]].

\bibitem{Gubler:2011ua} 
  P.~Gubler, K.~Morita and M.~Oka,
  Phys.\ Rev.\ Lett.\  {\bf 107}, 092003 (2011)
  [arXiv:1104.4436 [hep-ph]].

\bibitem{Walecka:1974qa} 
  J.~D.~Walecka,
  Annals Phys.\  {\bf 83}, 491 (1974).

\bibitem{Walecka:2004}
J. D. Walecka, in {\it Theoretical Nuclear and Subnuclear Physics}, 2nd ed. (World Scientific,Singapore, 2004), Chap. 18.

\bibitem{Tsushima:1998ru} 
  K.~Tsushima, D.~-H.~Lu, A.~W.~Thomas, K.~Saito and R.~H.~Landau,
  Phys.\ Rev.\ C {\bf 59}, 2824 (1999)
  [nucl-th/9810016].

\bibitem{Yasui:2012rw} 
  S.~Yasui and K.~Sudoh,
  arXiv:1207.3134 [hep-ph].

\bibitem{Kumar:2009xc} 
  A.~Kumar and A.~Mishra,
  arXiv:0912.2477 [nucl-th].

\bibitem{Morita:2010pd} 
  K.~Morita and S.~H.~Lee,
  Phys.\ Rev.\ C {\bf 85}, 044917 (2012)
  [arXiv:1012.3110 [hep-ph]].


\bibitem{Ghosh:2000zj} 
  S.~K.~Ghosh and B.~Jennings,
  Phys.\ Rev.\ C {\bf 61}, 067604 (2000)
  [nucl-th/0001047].

\bibitem{Gazda:2012zz}
  D.~Gazda and J.~Mares,
  Nucl.\ Phys.\ A {\bf 881} (2012) 159
  [arXiv:1206.0223 [nucl-th]].


\bibitem{Ding:2008gr} 
  G.~-J.~Ding,
  Phys.\ Rev.\ D {\bf 79}, 014001 (2009)
  [arXiv:0809.4818 [hep-ph]].
    
\bibitem{EE} E. Elizalde, {\it Ten physical applications of spectral
$zeta$ function}, Lecture Notes in Physics, Springer-Verlag, Berlin (1995).

\bibitem{Inagaki:1994ec}
  T.~Inagaki, T.~Kouno and T.~Muta,
  Int.\ J.\ Mod.\ Phys.\ A {\bf 10} (1995) 2241
  [hep-ph/9409413].


\bibitem{Abreu:2009zz} 
  L.~M.~Abreu, A.~P.~C.~Malbouisson, J.~M.~C.~Malbouisson and A.~E.~Santana,
  Nucl.\ Phys.\ B {\bf 819}, 127 (2009)
  [arXiv:0909.5105 [hep-th]].

\bibitem{Abreu:2011rj} 
  L.~M.~Abreu, A.~P.~C.~Malbouisson and J.~M.~C.~Malbouisson,
  Phys.\ Rev.\ D {\bf 83}, 025001 (2011)
  [arXiv:1102.1860 [hep-th]].
  

\bibitem{Padmanabhan:2004av} 
  T.~Padmanabhan,
  Curr.\ Sci.\  {\bf 88}, 1057 (2005)
  [astro-ph/0411044].

\bibitem{Albrecht:2006um} 
  A.~Albrecht, G.~Bernstein, R.~Cahn, W.~L.~Freedman, J.~Hewitt, W.~Hu, J.~Huth and M.~Kamionkowski {\it et al.},
  astro-ph/0609591.

\bibitem{Jakovac:2006sf} 
  A.~Jakovac, P.~Petreczky, K.~Petrov and A.~Velytsky,
  Phys.\ Rev.\ D {\bf 75}, 014506 (2007)
  [hep-lat/0611017].

\bibitem{Emerick:2011xu} 
  A.~Emerick, X.~Zhao and R.~Rapp,
  Eur.\ Phys.\ J.\ A {\bf 48}, 72 (2012)
  [arXiv:1111.6537 [hep-ph]].

\bibitem{Rapp:2008tf} 
  R.~Rapp, D.~Blaschke and P.~Crochet,
  Prog.\ Part.\ Nucl.\ Phys.\  {\bf 65}, 209 (2010)
  [arXiv:0807.2470 [hep-ph]].

\bibitem{Barosi:2008gx} 
  L.~Barosi, F.~A.~Brito and A.~R.~Queiroz,
  JCAP {\bf 0804}, 005 (2008)
  [arXiv:0801.0810 [hep-th]].

\end{references}
\end{document}